\documentclass[11pt,a4paper]{article}
\usepackage{graphicx}
\usepackage{amssymb}
\usepackage{amsmath}
\usepackage{amsfonts}
\usepackage{dsfont}
\usepackage{mathtools}
\usepackage{array}
\usepackage{rotating}
\usepackage{bbold,amsfonts}
\allowdisplaybreaks

\usepackage[utf8]{inputenc}
\usepackage{bm}
\usepackage{xcolor}
\usepackage{float}
\usepackage{braket}
\usepackage{placeins}
\usepackage[height=8.8in,width=6.45in]{geometry}
\usepackage[font=small,labelfont=bf]{caption}
\usepackage[hidelinks]{hyperref}
\usepackage{booktabs,float,slashed}
\usepackage{standalone}
\usepackage{caption}
\usepackage{subcaption}
\usepackage{makecell}
\numberwithin{equation}{section}
\newcommand{\vev}[1]{{\left\langle #1 \right\rangle}}

\newcommand{\beq}{\begin{equation}}
\newcommand{\eeq}{\end{equation}}

\newcommand{\overbar}[1]{\mkern 1.5mu\overline{\mkern-1.5mu#1\mkern-1.5mu}\mkern 1.5mu}

\DeclareMathOperator{\tr}{tr}

\newcommand{\ii}{\mathrm{i}}

\makeatletter
\newcommand*{\letterdef@}{}
\newcommand*{\letterdef}[3]{%
	\def\letterdef@##1{\expandafter\newcommand\csname #1\endcsname{#2{##1}}}%
	\@tfor\@tempa :=#3\do{\expandafter\letterdef@\expandafter{\@tempa}}}
\makeatother
\letterdef{c#1} {\mathcal}{ABCDEFGHIJKLMNOPQRSTUVWXYZ} 
\letterdef{rm#1}{\mathrm} {dDeimM} 

\DeclareMathAlphabet{\mathbb}{U}{msb}{m}{n} 

\DeclareMathAlphabet{\mathbb}{U}{msb}{m}{n} 

\begin{document}
\begin{titlepage}
\vspace*{10mm}
\begin{center}
{\LARGE \bf 
A note on integrated correlators \\[2mm]with a Wilson line
in $\mathcal{N}=4$ SYM
}

\vspace*{15mm}

{\Large M. Bill\`o${}^{\,a,b}$, M. Frau${}^{\,a,b}$, F. Galvagno${}^{\,c}$, A. Lerda${}^{\,d,b}$}

\vspace*{8mm}
	
${}^a$ Universit\`a di Torino, Dipartimento di Fisica,\\
			Via P. Giuria 1, I-10125 Torino, Italy
			\vskip 0.3cm
			
${}^b$   I.N.F.N. - sezione di Torino,\\
			Via P. Giuria 1, I-10125 Torino, Italy 
			\vskip 0.3cm
   
${}^c$  Centre for Theoretical Physics, Department of Physics and Astronomy,\\
Queen Mary University of London, London, E1 4NS, UK
			\vskip 0.3cm
   
${}^d$  Universit\`a del Piemonte Orientale,\\
			Dipartimento di Scienze e Innovazione Tecnologica\\
			Viale T. Michel 11, I-15121 Alessandria, Italy

\vskip 0.8cm
	{\small
		E-mail:
		\texttt{billo,frau,lerda@to.infn.it; f.galvagno@qmul.ac.uk}
	}
\vspace*{0.8cm}
\end{center}

\begin{abstract}
We revisit the analysis of the integrated 2-point functions of local operators with a $\frac{1}{2}$-BPS Wilson line in $\cN=4$ SYM. After including suitable parity-odd terms in the parametrization of the defect correlators, we are able to solve the superconformal Ward identities in terms of an unconstrained function of the cross-ratios. Exploiting this general solution, we obtain a simple expression of the integration measure for the integrated correlators with a Wilson line. We test our result by integrating the available bootstrap expression of the unintegrated correlator at strong coupling against the predictions of supersymmetric localization, finding perfect agreement.
\end{abstract}
\end{titlepage}
\setcounter{tocdepth}{2}
\tableofcontents

\section{Introduction}
\label{sec:intro}
The study of integrated correlators of local operators in superconformal gauge theories has recently attracted a lot of interest from several different points of view \cite{Binder:2019jwn,Chester:2019jas,Chester:2020dja,Chester:2020vyz,Dorigoni:2021bvj,Dorigoni:2021guq} being an ideal tool to explore non-perturbative physics
(see also \cite{Dorigoni:2022iem} for a review with all the relevant references).
A very promising recent development in this context concerns the study of integrated correlators in $\cN=4$ SU($N$) SYM in presence of a line defect, like for example a $\frac{1}{2}$-BPS Wilson line \cite{Pufu:2023vwo,Billo:2023ncz}. This new class of integrated correlators can be obtained by considering a Wilson line $W$ in the mass-deformed $\cN=2^*$ theory and taking two mass-derivatives of the logarithm of its vacuum expectation value according to
\begin{equation}
\label{eq:1.2}
    \int\!d^4 x_1 d^4 x_2 ~~ \widehat{\mu}(x_1,x_2)~ \big\langle O_2(x_1) O_2(x_2)\big\rangle_W = \partial^2_{m} \log \big\langle W\big\rangle_{\mathcal{N}=2^*}\Big|_{m=0}~.
\end{equation}
Here $O_2(x_i)$ are scalar primary operators of conformal dimension 2 in the $\mathbf{20^\prime}$ representation of the $R$-symmetry group of the $\cN=4$ SYM, whose 2-point functions in presence of a Wilson line we have denoted by $\big\langle O_2(x_1) O_2(x_2)\big\rangle_W$, 
and $\widehat{\mu}(x_1,x_2)$ is an integration measure, which is the main focus of this note. On the other hand, the vacuum expectation value of the Wilson line in the massive theory, {\it{i.e.}} $\vev{W}_{\cN=2^*}$, can be computed explicitly with matrix-model techniques using supersymmetric localization \cite{Pestun:2007rz,Buchel:2013id,Russo:2013kea}. Therefore, (\ref{eq:1.2}) can be regarded as an integral constraint that localization puts on the gauge theory correlators in presence of a line defect.

The relation (\ref{eq:1.2}) was first considered in \cite{Pufu:2023vwo}, where the right-hand side has been studied in depth by computing the matrix model in the large-$N$ expansion, including also several corrections beyond the planar level. The left-hand side of (\ref{eq:1.2}) has been studied in \cite{Billo:2023ncz}, focusing on the derivation of the integration measure $\widehat{\mu}$. This derivation was based on a
solution of the superconformal Ward identities obeyed by the 2-point functions in presence of a Wilson line which was obtained by considering the structures of the defect CFT correlators discussed in \cite{Lauria:2018klo,Herzog:2020bqw}. The solution found in this way, however, constrains in a severe way the functions of the conformal cross-ratios that appear in the correlation functions. In this note we reconsider this issue and show that by including new structures (which are parity odd)\,%
\footnote{We are deeply indebted to Lorenzo Bianchi and Maddalena Ferragatta who raised this point with us and prompted us to reconsider our original solution and also for re-deriving and checking the system of Ward identities.}, it is possible to solve the Ward identities in terms of an unconstrained function of the cross-ratios and its derivatives.
Exploiting this new solution, we obtain a very simple expression of the integration measure $\widehat{\mu}$, which amends the one derived in \cite{Billo:2023ncz} and is presented in Section~\ref{sec:measure} (see in particular (\ref{Afin}) and (\ref{Afin2})).

As mentioned above, the right-hand side of (\ref{eq:1.2}) can be efficiently computed using matrix-model techniques which, in the large-$N$ limit, allow us to obtain explicit expressions valid for all values of the 't Hooft coupling. These matrix-model results can be used to check the validity of the left-hand side, by integrating the existing results for the un-integrated two-point correlator.
In \cite{Barrat:2021yvp} (see also \cite{Gimenez-Grau:2023fcy}) a general expression for $\big\langle O_2(x_1) O_2(x_2)\big\rangle_W$ at strong-coupling has been derived in the planar limit using bootstrap methods and thus this result can be used to check the validity of (\ref{eq:1.2}) at least in this regime.
As we show in Section~\ref{sec:matrix}, after integrating the strong-coupling expression of the
un-integrated correlator given in \cite{Barrat:2021yvp} over the measure we have derived, we find a perfect match with the matrix-model
result that follows from supersymmetric localization. We regard this successful check as a very convincing sign of the correctness of our results. Similar checks are not yet fully available at weak coupling, but this final form for the integration measure can then be used as a constraint for Feynman diagram computations \cite{Barrat:2020vch} and for further bootstrap computations.

\vspace{0.7cm}
{\textbf{Note added}}: While this note was in preparation, we became aware that similar results have been obtained in \cite{DOPW}. We thank the authors of \cite{DOPW} for sharing a draft of their paper prior to publication. Despite the different formalisms used, both our results agree with each other.

\section{Integrated correlators with a Wilson line}
\label{sec:integrated}
The presentation in this Section closely follows \cite{Billo:2023ncz}
whose notations and conventions we adopt and to which we refer for details.

We consider a half-BPS circular Wilson loop $W$ in $\mathcal{N}=2^*$ SYM on a 4-sphere $S^4$ in the fundamental representation of SU($N$):
\begin{equation}
\label{defWCsusy}
W=\frac{1}{N}\tr \, \cP
		\exp \bigg\{\sqrt{\frac{\lambda}{N}} \oint_{\cC} d\tau \Big[\ii \,A_\mu(x)\,\dot x^{\mu}(\tau)
+\frac{\varphi(x)+\overbar{\varphi}(x)}{\sqrt{2}}\Big] \bigg\}~.
\end{equation}
Here $\cC$ is the equator of $S^4$, $\lambda = g_{\rm{YM}}^2 N$ is the 't Hooft coupling, and 
the scalars $\varphi$ and $\overbar{\varphi}$ belong to the $\cN=2$ vector multiplet together with the gauge vector $A_\mu$.

The main quantity of interest is the vacuum expectation value of $W$ in the massive theory, namely
\begin{equation}
    \big\langle \!\! \big\langle W \big\rangle \!\! \big\rangle_{\cN=2^*} :=\,
    \frac{\displaystyle{\int \cD[\mathrm{fields}]~W~
    \rme^{-S_{\cN=4}-S_m}}}{\displaystyle{\int \cD[\mathrm{fields}]\,\,
    \rme^{-S_{\cN=4}-S_m}}}
    \label{vevW}
\end{equation}
where $S_{\cN=4}$ is the usual action of $\cN=4$ SYM on $S^4$ \cite{Binder:2019jwn,Chester:2020dja,Pufu:2023vwo} and $S_m$ is the mass-deformation 
\begin{equation}
    S_m=m\int\!d^4x\,\sqrt{g(x)}\,
    \big(\ii\,J(x)+K(x)\big)~.
    \label{Sm1}
\end{equation}
Here $g(x)$ is the determinant of the metric of $S^4$ (whose radius we have set to 1) while 
$J(x)$ and $K(x)$ are conformal primaries with dimensions 2 and 3 respectively that can be written in terms of the components of the so-called $\mathcal{N}=2$ current multiplet (see for example
\cite{Dolan:2001tt})\,%
\footnote{Notice that the action of the $\cN=2^*$ theory also contains a term proportional to $m^2$ but, as argued in \cite{Binder:2019jwn,Chester:2020dja,Dorigoni:2021guq,Pufu:2023vwo}, it does not play
any significant role since its contributions cancel against boundary terms that are generated by integration by parts. Neglecting this term means that in practice one can freely perform integrations by parts.}. The field $J$ is usually known as moment-map operator.

From (\ref{vevW}), it easily follows that
\begin{equation}
\begin{aligned}
    \mathcal{I} := 
   \partial_m^2 \log \big\langle \!\! \big\langle W\big\rangle\!\!\big\rangle_{\cN=2^*} &\Big|_{m=0}
    =-\int\!d^4x_1\,\sqrt{g(x_1)}
    \int\!d^4x_2\,\sqrt{g(x_2)}~\frac{\big\langle \!\!\big\langle
    W\,J(x_1)\,J(x_2)\big\rangle\!\!\big\rangle^{c}}{\big\langle\!\!\big\langle W \big\rangle\!\!\big\rangle}\\[2mm]
    &+\int\!d^4x_1\,\sqrt{g(x_1)}
    \int\!d^4x_2\,\sqrt{g(x_2)}~\frac{\big\langle\!\!\big\langle
    W\,K(x_1)\,K(x_2)\big\rangle\!\!\big\rangle^{c}}{\big\langle\!\!\big\langle W \big\rangle\!\!\big\rangle}
    \label{Iis2}
\end{aligned}
\end{equation}
where the superscript $c$ stands for connected. 
Using a stereographic projection from $S^4$ to $\mathbb{R}^4$ with a conformal factor given by $\Omega(x)=(1+x^2)/2$ under which the circular Wilson loop (\ref{defWCsusy}) is mapped to a straight Wilson line in $\mathbb{R}^4$, one can rewrite the sphere correlators in terms of the flat-space correlators using the following relation
\begin{equation}
    \big\langle\!\!\big\langle
    W~\cO(x_1)\,\cO(x_2)\big\rangle\!\!\big\rangle^c
    =\bigg(\frac{1+x_1^2}{2}\bigg)^\Delta\,\bigg(\frac{1+x_2^2}{2}\bigg)^\Delta\,
    \big\langle
    \cO(x_1)\,\cO(x_2)\big\rangle_W^c
\end{equation}
where $\Delta$ is the conformal dimension of $\mathcal O(x_i)$
(see \cite{Billo:2023ncz} for details).
Proceeding in this way, we obtain
\begin{equation}
\begin{aligned}
   \mathcal{I}&=-\int\!d^4x_1\!\int\!d^4x_2\,\frac{1}{\displaystyle{\bigg(\frac{1+x_1^2}{2}\bigg)^2\bigg(\frac{1+x_2^2}{2}\bigg)^2}}
    \,\,\big\langle
    J(x_1)\,J(x_2)\big\rangle_{W}\\[2mm]
    &\qquad+\int\!d^4x_1\!\int\!d^4x_2\,\frac{1}{\displaystyle{\bigg(\frac{1+x_1^2}{2}\bigg)
    \bigg(\frac{1+x_2^2}{2}\bigg)}}
    \,\,\big\langle
    K(x_1)\,K(x_2)\big\rangle_{W}
    \label{Iis3}
\end{aligned}
\end{equation}
where we have dropped the superscript $c$ for simplicity and used the fact that in flat space $\langle W \rangle=1$ if the Wilson line is normalized as in (\ref{defWCsusy}). 

Conformal symmetry in presence of a defect is broken to SO$(1,2) \times $ SO$(3)$ and severely constrains the form of the correlators of primary fields \cite{Billo:2016cpy}, which in our case are
\begin{align}
    \big\langle
    J(x_1)\,J(x_2)\big\rangle_{W} = \frac{F(\xi,\eta)}{r_1^2\,r_2^2} ~,~~~~
    \big\langle K(x_1)\,K(x_2)\big\rangle_{W}=\frac{G(\xi,\eta)}{r_1^3\,r_2^3}~.
    \label{WJJWKK}
\end{align}
Here $r_i=|\Vec{x}_i|$ represents the orthogonal distance between the point $x_i$ and the line, and $F$ and $G$
are functions of the conformal invariant cross-ratios $\xi$ and $\eta$ defined as \cite{Buchbinder:2012vr}
\begin{equation}
    \xi=\frac{(x_{12}^4)^2+r_1^2+
    r_2^2}{2\,r_1\,r_2}~,\qquad
    \eta=\frac{\vec{x}_1\cdot \vec{x}_2}{r_1\,r_2}~,
  \label{eq:CrossRatios}
\end{equation}
where $x_{12}^4=(x_1^4-x_2^4)$, with $x_i^4$ being the longitudinal coordinate along the line.

Exploiting the SO(3) symmetry, we can rotate the transverse axes so that the points $x_1$ and $x_2$ take the form
\begin{align}
    x_1=(r_1,0,0,t_1)~~~\mbox{and}~~~ x_2=(r_2\cos \theta,r_2\sin\theta,0,t_2)
    \label{x1x2}
\end{align}
and the cross-ratios (\ref{eq:CrossRatios}) become
\begin{align}
    \xi=\frac{(x_{12}^4)^2+r_1^2+r_2^2}{2\,r_1 r_2}~,~~~
    \eta=\cos\theta~.
    \label{xieta1}
\end{align}
Exploiting the SO(1,2) symmetry, we can bring the points $x_1$ and $x_2$ in \eqref{x1x2} to the reference configuration
\begin{align}
     x_1=(1,0,0,0)~~~\mbox{and}~~~ x_2=(\rho\cos \theta,\rho\sin\theta,0,0)
    \label{x1x2bis}
\end{align}
in which the invariant $\xi$ becomes
\begin{align}
    \xi=\frac{1}{2}\Big(\rho+\frac{1}{\rho}\Big)~,
    \label{xirho}
\end{align}
while $\eta$ remains equal to $\cos\theta$. By trading some of the integrals of (\ref{Iis3}) for the integrals over the parameters of the transformations that are used to reach the reference configuration, in \cite{Billo:2023ncz} it has been shown that the integrated correlator $\cI$ can be recast in the following form
\begin{align}
    \mathcal{I}=128\pi^4\!\!\int_{-1}^{+1}\!\!d\eta\int_0^1\!\!d\rho\,\bigg[\frac{\big(1-\rho^2+(1+\rho^2) 
    \log\rho\big)}{(1-\rho^2)^2}\,F(\xi,\eta)-\frac{
    \log\rho}{2\rho}\,G(\xi,\eta)\bigg]_{\xi=\frac{1}{2}(\rho+\frac{1}{\rho})}~.
    \label{Ifin}
\end{align}
To proceed we need to find a relation between the functions $F$ and $G$. This can be done by exploiting the supersymmetric Ward identities satisfied by the bulk correlators in presence of a Wilson line.

\section{Ward identities and their solution}
Before discussing the Ward identities, we recall that fields $J$ and $K$, whose 2-point functions
appear in the integrated correlator (\ref{Ifin}), are built from the so-called $\cN=2$ current multiplet
which consists of three scalars $\Phi^{ij}=\Phi^{ji}$, such that
$(\Phi^{ij})^*=\epsilon_{ik}\,\epsilon_{j\ell}\,\Phi^{k\ell}$, two chiral fermions $X_{i}$,
two anti-chiral fermions $\overbar{X}_i$, two real scalars $P$ and $\overbar{P}$, and one conserved current $j_\mu$\,%
\footnote{For our conventions on indices and spinors we refer to Appendix~A of \cite{Billo:2023ncz}). Here we simply recall that $i,j = 1,2$ are SU(2)$_R$ symmetry indices.}. In presence of a line defect, the original rotation symmetry SO(4) of $\mathbb{R}^4$ is reduced to SO(3) and thus it is convenient to repackage these fields in the following combinations
\begin{align}
    \Phi^{ij}~,~~Y_{i}=X_{i}+ \ii\,\overbar{X}_i~,~~ Z_{i}=X_{i} - \ii\,\overbar{X}_i~,~~
    \vec{j}~,~~ S=Q-2\,j^4~,~~T~,
    \label{newcurrmult}
\end{align}
where
\begin{equation}
    T=P+\overbar{P}~, ~~~~Q=P-\overbar{P}~.
    \label{QTvsPPbar}
\end{equation}
The advantage of using these combinations is that the supersymmetry transformations preserved by the Wilson line take a simple form as shown below:
\begin{subequations}
\begin{align}
\delta\Phi^{ij} & = -\frac{1}{2}\xi^{i}\,Y^{j}+\left(i\leftrightarrow j\right)~,\\
\delta Y^i & = S\,\xi^i - 4 \,\ii\, \partial_4 \Phi^{ij}\,\xi_j~,\\
\delta S & = 2\,\ii\,\xi^i \,\partial_4 Y_i~,\\
\delta Z^i & = T\,\xi^i - 2\,\ii\, \vec j\cdot\vec\sigma\,\xi^i 
+ 4\,\vec\partial\Phi^{ij} \cdot \vec\sigma\,\xi_j~,\\
\delta T & = 2\,\ii\,\xi^i\,\partial_4Z_i + 2\xi^i\,\vec\sigma\cdot\vec\partial Y_i~,\\
\delta \vec j & = - \xi^i\, \vec\sigma\, \partial_4 Z_i
+ \xi^i \,\vec\sigma\wedge \vec\partial Y_i~.
\end{align}
\label{SUSYcurrentBPS}%
\end{subequations}

Another ingredient is the fact that residual SU(2)$_R$ $R$-symmetry and the residual conformal symmetry constrain 
the 2-point function of $\Phi^{ij}$ 
(which is a conformal primary operator of dimension 2) to have the following form
\begin{align}       
    \label{formcorr2}
        \big\langle \Phi^{i j}(x_1)\,\Phi^{k \ell}(x_2)\big\rangle_W 
        & = \left(\epsilon^{i k}\,\epsilon^{j \ell} + \epsilon^{i \ell}\,\epsilon^{j k}\right) A(x_1,x_2)~,
\end{align}
where $A(x_1,x_2)$ is a symmetric function of its arguments fixed by conformal symmetry to be:
\begin{align}
   A(x_1,x_2)= \frac{a(\xi,\eta)}{r_1^2\,r_2^2}
   \label{Avsa}
\end{align}
where $a$ is a function of the invariant cross-ratios (\ref{eq:CrossRatios}).
Furthermore, the correlators of $\Phi^{ij}$ with the other bosonic operators in the multiplet vanish:
\begin{align}
    \label{formcorr1}
       \big\langle \Phi^{ij}(x_1)\,S(x_2)\big\rangle_W & = \left\langle \Phi^{ij}(x_1)\,T(x_2)\right\rangle_W 
       = \left\langle \Phi^{ij}(x_1)\,\vec j(x_2)\right\rangle_W = \left\langle \Phi^{ij}(x_1)\,j^4(x_2)\right\rangle_W = 0~.
\end{align}
This is because $\Phi^{ij}$ is symmetric in $i$ and $j$ and the only $R$-invariant 2-index tensor at our disposal is the anti-symmetric $\epsilon^{ij}$. 
Similar $R$-symmetry considerations lead us to write 
\begin{subequations}
    \begin{align}
        \big\langle Y^i_\alpha(x_1)\,Y^j_\beta(x_2)\big\rangle_W 
        & = \epsilon^{ij}\left(\epsilon_{\alpha\beta}\, b(x_1,x_2) + \vec\tau_{\alpha\beta}\cdot \vec b(x_1,x_2)\right)~,\\[1.2mm]
        \big\langle Z^i_\alpha(x_1)\,Y^j_\beta(x_2)\big\rangle_W 
        & = \epsilon^{ij}\left(\epsilon_{\alpha\beta}\, c(x_1,x_2) + \vec\tau_{\alpha\beta}\cdot \vec c(x_1,x_2)\right)~,\\[1.2mm]
        \big\langle Z^i_\alpha(x_1)\,Z^j_\beta(x_2)\big\rangle_W 
        & = \epsilon^{ij}\left(\epsilon_{\alpha\beta}\, d(x_1,x_2) + \vec\tau_{\alpha\beta}\cdot \vec d(x_1,x_2)\right)~.
\end{align}
 \label{formcorr3}%
\end{subequations}
where $\Vec{\tau}_{\alpha\beta} \equiv \vec{\sigma}_{\alpha}^{~\gamma}\,\epsilon_{\gamma\beta}$ is symmetric in $\alpha$ and $\beta$. The functions $b(x_1,x_2)$ and $d(x_1,x_2)$ are anti-symmetric in their arguments, while $\vec b(x_1,x_2)$ and $\vec d(x_1,x_2)$ are symmetric. Finally, taking into account that $T$ and $Q$ are conformal scalar primaries with dimension 3, we have
\begin{subequations}\label{eq:TT_QQ}
    \begin{align}
\big\langle T(x_1)\,T(x_2)\big\rangle_W=& \,H(x_1,x_2)=\frac{h(\xi,\eta)}{r_1^3\,r_2^3}~,\label{TT}\\
 \big\langle Q(x_1)\,Q(x_2)\big\rangle_W=&\,L(x_1,x_2)=\frac{\ell(\xi,\eta)}{r_1^3\,r_2^3}~.\label{QQ}
\end{align}
\end{subequations}
Also the correlator of two currents in a defect CFT depends on the coordinates and the cross-ratios
in a very specific form which has been derived in \cite{Lauria:2018klo,Herzog:2020bqw}.
In particular, following the parametrization of \cite{Herzog:2020bqw}
we have
\begin{subequations}\label{eq:jj_general}
    \begin{align}
    \left\langle j^m(x_1)\,j^n(x_2)\right\rangle_W &=\frac{1}{r_1^3 r_2^3}\bigg[\delta^{mn}\bigg(\frac{f_4+f_5}{64u^3}\bigg)
    +\frac{x_1^mx_2^n}{r_1r_2}\bigg(\frac{f_4}{128u^4}+\frac{(1+4u^2)f_1}{256u^5}+\frac{f_2+2f_3}{64u^3}\bigg)\notag\\[1mm]
    &\qquad~~~~+\frac{x_1^nx_2^m}{r_1r_2}\bigg(\frac{f_4}{128u^4}-\frac{f_5}{64u^3 \eta}+\frac{f_1}{256u^5}+
\frac{f_2}{64u^3 \eta^2}-\frac{f_3}{64u^4 \eta}\bigg)\notag\\[1mm]
    &\qquad~~~~+\frac{x_1^mx_1^n+x_2^mx_2^n}{r_1 r_2}\bigg(\frac{f_3}{128u^4  \eta}-\frac{f_1}{256u^5}-\frac{f_4}{128u^4}\bigg)\notag\\[1mm]
    &\qquad~~~~+\bigg(\frac{x_1^mx_2^n}{r_1^2}+\frac{x_1^mx_2^n}{r_2^2}\bigg)\bigg(\!\!-\frac{f_1+f_3}{128u^4}\bigg)\label{jmjn}\\[1mm]
    &\qquad~~~~+\bigg(\frac{x_1^mx_1^n}{r_1^2}+\frac{x_2^mx_2^n}{r_2^2}\bigg)\bigg(\frac{f_1}{128u^4}-\frac{f_2}{64u^3\eta}+\frac{(\eta-2u)f_3}{128u^4\eta}\bigg)\bigg]~,\notag\\[2mm]
     \left\langle j^m(x_1)\,j^4(x_2)\right\rangle_W&=-\frac{x_{12}^4}{r_1^3r_2^3}\bigg[\frac{x_1^m}{r_1 r_2}\bigg(\frac{f_1}{256u^5}+\frac{f_4}{128u^4}\bigg)+\frac{x_2^m}{r_1 r_2}\bigg(\frac{f_3}{128u^4\eta}-\frac{f_1}{256u^5}-\frac{f_4}{128u^4}\bigg)\notag\\[1mm]
&\qquad~~~~~~+ \frac{x_1^m}{r_1^2}\bigg(\!\!-\frac{f_1+f_3}{128u^4}\bigg)\bigg]~,\label{jmj4}\\[2mm]
 \left\langle j^4(x_1)\,j^4(x_2)\right\rangle_W&=\frac{1}{r_1^3\,r_2^3}
 \bigg[\frac{f_4}{64u^3}-\frac{(x_{12}^4)^2}{r_1r_2}\bigg(\frac{f_1}{256\,u^5}+\frac{f_4}{128\,u^4}\bigg)\bigg]~,\label{j4j4}
\end{align}
\label{jj}%
\end{subequations}
where $f_1,\ldots, f_5$ are five functions of the invariants and we have defined $u=(\xi-\eta)/2$ for convenience.

The last ingredient we need is the form of the correlators between the current and the scalars $T$ and $Q$, which are the novelty compared to \cite{Billo:2023ncz}. The current/scalar correlators have also been discussed in \cite{Lauria:2018klo,Herzog:2020bqw}, and in our case, using again the parametrization of \cite{Herzog:2020bqw}, they are
\begin{subequations}\label{eq:jT_general}
    \begin{align}
        \big\langle j^m(x_1)\,T(x_2)\big\rangle_W&=\frac{1}{r_1^3r_2^3}\bigg[\!-\frac{x_1^m}{r_1}(g_1+g_2)
    +\frac{x_2^m}{r_2}\Big(\frac{g_2}{\eta}-\frac{g_1}{2u}\Big)+\frac{x_1^m}{r_2}\Big(\frac{g_2}{2u}
    \Big)\bigg]\label{jmTeven}~,\\[2mm]
     \big\langle j^4(x_1)\,T(x_2)\big\rangle_W&=\frac{1}{r_1^3r_2^3}\bigg[\frac{x_{12}^4}{r_2}\Big(\frac{g_1}{2u}\Big)\bigg]~,
    \end{align}
\end{subequations}
and
\begin{subequations}\label{eq:jQ_general}
    \begin{align}
        \big\langle j^m(x_1)\,Q(x_2)\big\rangle_W&=\frac{1}{r_1^3r_2^3}\bigg[\!-\frac{x_1^m}{r_1}(g^\prime_1+g^\prime_2)
    +\frac{x_2^m}{r_2}\Big(\frac{g^\prime_2}{\eta}-\frac{g^\prime_1}{2u}\Big)+\frac{x_1^m}{r_2}\Big(\frac{g^\prime_2}{2u}
    \Big)\bigg]~,\\[2mm]
    \big\langle j^4(x_1)\,Q(x_2)\big\rangle_W&=\frac{1}{r_1^3r_2^3}\bigg[\frac{x_{12}^4}{r_2}\Big(\frac{g^\prime_1}{2u}\Big)\bigg]
    \end{align}
\end{subequations}
where $g_1$, $g_2$, $g^\prime_1$ and $g^\prime_2$ are four functions of the invariants. However, these formulas are not sufficient to completely determine the current/scalar correlators. This is because the scalar operators $T$ and $Q$ have a specific
parity under the inversion along the line and in the transverse space. More precisely, $T$ is even under the longitudinal parity and odd under the transverse parity, while viceversa $Q$ is odd under the longitudinal parity and even under the transverse one. In view of this fact, further structures in the correlators are possible. Indeed, as shown in Appendix~\ref{app_odd} using the embedding formalism, one has
\begin{subequations}
    \begin{align}
\big\langle j^m(x_1) \,T(x_2)\big\rangle_W^{\mathrm{odd}} &= \frac{1}{r_1^3r_2^3}
\bigg[\frac{\epsilon^{mnp}\,x_1^mx_2^p}{r_1 r_2}\,s\bigg]~,\\[1mm]
~~\big\langle j_4 (x_1) \,T(x_2)\big\rangle_W^{\mathrm{odd}}&=0~,
\end{align}
\label{Todd}%
\end{subequations}
and
\begin{subequations}
    \begin{align}
\big\langle j^m(x_1) \,Q(x_2)\big\rangle_W^{\mathrm{odd}} &=\frac{1}{r_1^3r_2^3}
\bigg[\frac{x_{12}^4\,x^m_1}{r_1r_2}\,2s^\prime\bigg]~,\\[1mm]
~~\big\langle j_4 (x_1) \,Q(x_2)\big\rangle_W^{\mathrm{odd}}&=\frac{1}{r_1^3r_2^3}
\bigg[\frac{(x_{12}^4)^2}{r_1r_2}\,s^\prime-\frac{r_1}{r_2}\,s^\prime+\frac{r_2}{r_1}\,s^\prime\bigg]
\end{align}
\label{Qodd}%
\end{subequations}
where $s$ and $s^\prime$ are functions of the invariants. These odd structures are the new crucial ingredient in our analysis.

The task now is to find relations among the various functions of the invariants appearing in the previous formulas. This can be achieved by writing the identities that follow from applying a supersymmetry transformation to a vanishing boson/fermion correlator. These calculations have been described in great detail in \cite{Billo:2023ncz} (see in particular Appendix D). Therefore, we do not repeat them here and simply report the final results, which are
\begin{subequations}
    \begin{align}
    \big\langle Y^i_\alpha(x_1)\,Y^j_\beta(x_2)\big\rangle_W&=8\,\ii\,\epsilon^{ij}\,\epsilon_{\alpha\beta}\,\partial^{(2)}_4A(x_1,x_2)~,\\[1mm]
    \big\langle Z^i_\alpha(x_1)\,Y^j_\beta(x_2)\big\rangle_W&=8\,\epsilon^{ij}\,\tau^m_{\alpha\beta}\,\partial_m^{(1)}A(x_1,x_2)~,\\[1mm]
\big\langle S(x_1)\,S(x_2)\big\rangle_W&=-16\, \partial^{(1)}_4 \partial^{(2)}_4 A(x_1,x_2)~,\label{SS}\\[1mm]
\big\langle T(x_1)\,S(x_2)\big\rangle_W&=0~,\\[1mm]
\big\langle j^m(x_1)\,S(x_2)\big\rangle_W&=8\, \partial_m^{(1)}\,\partial^{(2)}_4 A(x_1,x_2)~,\label{jS}\\[1mm]
\big\langle T(x_1)\,T(x_2)\big\rangle_W&=-2\ii\,\partial^{(2)}_4 d(x_1,x_2)+16 \,\partial_m^{(2)}\,\partial_m^{(1)}A(x_1,x_2)~,\\[1mm]
\big\langle j^m(x_1)\,T(x_2)\big\rangle_W&=\partial^{(2)}_4 d^m(x_1,x_2)+8 \,\epsilon^{mnp}\,\partial_n^{(2)}\partial_p^{(1)}A(x_1,x_2)~,\label{jmT}\\[1mm]
\big\langle j^m(x_1)\,j^n(x_2)\big\rangle_W&=\frac 14 \big\langle T(x_1)\,T(x_2)\big\rangle_W\,\delta^{mn}-\frac{1}{2}\epsilon^{mnp} \big\langle j^p(x_1)\,T(x_2)\big\rangle_W- 4\, \partial^{(1)}_m\partial^{(2)}_n A(x_1,x_2)~.
\label{jjnew}
\end{align}
\label{Wardidentities}%
\end{subequations}
Inserting in these formulas the structure of the correlators prescribed by the defect CFT that we displayed above, we can find a set of equations that must be satisfied by the various functions of the invariants.

We note in particular that the relation (\ref{jmT}) implies that
\begin{align}
    \big\langle j^m(x_1)\,T(x_2)\big\rangle_W+
    \big\langle T(x_1) \,j^m(x_2)\big\rangle_W=0
    \label{jTTj}
\end{align}
and if we use the parametrization in (\ref{jmTeven}) one can easily conclude that the functions $g_1$ and $g_2$ vanish, thus implying the vanishing of the whole correlator. This was used in \cite{Billo:2023ncz} to drop the $\epsilon$-term in the last Ward identity (\ref{jjnew}). With such a choice, however, the system of equations descending from the Ward identities admits a solution only if one imposes a homogeneity constraint on the scalar function $a$ appearing in (\ref{Avsa}). On the other hand, the odd structure (\ref{Todd}) is perfectly compatible with the relation
(\ref{jTTj}) for any function $s$, which therefore remains as an available degree of freedom that allows us to lift that constraint. Similarly, the function $s^\prime$
appearing in the odd structures (\ref{Qodd}) will enter the identities (\ref{SS}) and (\ref{jS}), once we recall that $S=Q-2j^4$, thus modifying the corresponding relations.

Plugging the form of the correlators prescribed by $R$-symmetry and conformal symmetry in the Ward identities (\ref{Wardidentities}) and proceeding as described in \cite{Billo:2023ncz}, we obtain a system of linear equations for the various functions of the invariants that admits a solution. In this way we find that $g_1,g_2,g_1^\prime$ and $g_2^\prime$ vanish while the
functions $f_1,\cdots,f_5,s,s^\prime,h$ and $\ell$ are non trivial and can be written in terms of the function $a$ and its derivatives. 
In particular, we obtain the following expressions 
\begin{align}
   h&=16\,\partial_\xi\Big(\eta (1-\xi^2)\,\partial_\xi a\Big)+16 \,\partial_\eta\Big((\eta^2-1)a+\eta(\eta^2-1)\partial_\eta a\Big)~,
   \label{h}\\[2mm]
   \ell&=16\,\partial_\eta\Big(\xi (1-\eta^2)\,\partial_\eta a\Big)+16 \,\partial_\xi\Big((\xi^2-1)a+\xi(\xi^2-1)\partial_\xi a\Big)~,\label{ellfin}\\[2mm]
   s&=8 (\xi ^2-1)\partial_\xi^2 a-8(\eta ^2-1) \partial_\eta^2 a+16 \,\xi  \partial_\xi a-32 \,\eta  \partial_\eta a-16 a~,\\[2mm]
   s^\prime&=8 (\eta ^2-1)\partial_\eta^2 a-8 (\xi ^2 -1)\partial_\xi^2 a +16\, \eta  \partial_\eta a-32 \,\xi \partial_\xi a -16a~.
\end{align}
We observe a curious symmetry of these expressions under the exchange of $\xi$ and $\eta$.

\section{Final form of the integrated correlator}
\label{sec:measure}
We now have all ingredients to obtain the final form of the integrated correlator (\ref{Ifin}).
We first recall that since $J=\Phi^{11}+\Phi^{22}$,
from (\ref{formcorr2}) it follows that
\begin{align}
    \big\langle J(x_1)\,J(x_2)\big\rangle_W
    =\frac{4\,a(\xi,\eta)}{r_1^2\,r_2^2}~,
\end{align}
implying that
\begin{equation}
    F(\xi,\eta)=4\,a(\xi,\eta)~.
    \label{F_to_a}
\end{equation}
On the other hand, $K$ is a bi-linear combination of hypermultiplet fermions\,%
\footnote{We take this opportunity to correct a sign error in the expression of $\overbar{P}$ in terms of the anti-chiral fermions of the hypermultiplet that led in \cite{Billo:2023ncz} to write $K$ as proportional to $T=P+\overbar{P}$ instead of $Q=P-\overbar{P}$.}, which corresponds to $Q/2$. Thus, from (\ref{QQ}) we have
\begin{align}
    \big\langle K(x_1)\,K(x_2)\big\rangle_W=
    \frac{1}{4}\,\frac{\ell(\xi,\eta)}{r_1^3\,r_2^3}~,
\end{align}
yielding
\begin{equation}
    \begin{aligned}
    G(\xi,\eta)&=\frac{1}{4}\,\ell(\xi,\eta)
    \\&
    =\,\partial_\eta\Big[\xi (1-\eta^2)\,\partial_\eta F(\xi,\eta)\Big]+\partial_\xi\Big[(\xi^2-1)F(\xi,\eta)+\xi(\xi^2-1)\partial_\xi F(\xi,\eta)\Big]
\end{aligned}
 \label{G_to_F}
\end{equation}
where in the second step we used the solution of the Ward identities
given in (\ref{ellfin}) and used (\ref{F_to_a}).

Since the integration variable $\rho$ used in (\ref{Ifin}) is only a function of $\xi$ (see (\ref{xirho})) and the integration measure is $\eta$-independent, we can drop the total $\eta$-derivative and use for $G$ the following reduced expression
\begin{equation}
    G(\xi,\eta)=\partial_\xi\Big[(\xi^2-1)F(\xi,\eta)+\xi(\xi^2-1)\partial_\xi F(\xi,\eta)\Big]
    ~.
    \label{G_to_F1}
\end{equation}
Upon inserting this into (\ref{Ifin}) we get
\begin{align}
    \mathcal{I}&= 
   128\pi^4\!\!\int_0^1\!\!d\rho\int_{-1}^{+1}\!\!d\eta\,\bigg\{\frac{\big(1-\rho^2+(1+\rho^2) 
    \log\rho\big)}{(1-\rho^2)^2}\,F(\xi,\eta)\notag\\
    &\qquad\qquad\qquad\qquad-\frac{
    \log\rho}{2\rho}\,\partial_\xi\Big[(\xi^2-1)F(\xi,\eta)+\xi(\xi^2-1)\partial_\xi F(\xi,\eta)\Big]\bigg\}_{\xi=\frac{1}{2}(\rho+\frac{1}{\rho})}~.
    \label{Ifin2}
\end{align}
Performing the change of variable from $\xi$ to $\rho$, we find
\begin{align}
    \mathcal{I}&= 
   128\pi^4\!\!\int_0^1\!\!d\rho\int_{-1}^{+1}\!\!d\eta\,\bigg\{\frac{\big(1-\rho^2+(1+\rho^2) 
    \log\rho\big)}{(1-\rho^2)^2}\,\widehat{F}(\rho,\eta)\notag\\
    &\qquad\qquad\qquad\qquad-\frac{
    \rho \log\rho}{\rho^2-1}\,\partial_\rho\Big[\frac{(\rho^2-1)^2}{4\rho^2}\widehat{F}(\rho,\eta)+\frac{\rho^4-1}{4\rho}\partial_\rho \widehat{F}(\rho,\eta)\Big]\bigg\}
    \label{Ifin3}
\end{align}
where
\begin{align}
    \widehat{F}(\rho,\eta)\,\equiv\,F(\xi,\eta)\bigg|_{\xi=\frac{1}{2}\big(\rho+\frac{1}{\rho}\big)}~.
\end{align}
If in the second line of (\ref{Ifin3}) we integrate by parts (dropping the boundary terms), we find a remarkable simplification
between the various pieces and the final form of the integrated correlator is simply
\begin{align}
    \mathcal{I}&=128\pi^4\!\!\int_0^1\!\!d\rho\int_{-1}^{+1}\!\!d\eta\,\,\frac{\rho^2-1}{4\rho^2}\,\widehat{F}(\rho,\eta)~.
    \label{Afin}
\end{align}
Using $\xi$ instead of $\rho$ as integration variable, we can rewrite $\cI$ as
\begin{align}
    \mathcal{I}&=-64\pi^4\!\!\int_1^\infty\!\!d\xi\int_{-1}^{+1}\!\!d\eta\,\,F(\xi,\eta)~.
    \label{Afin2}
\end{align}
This is the main result of this note.

\section{The matrix-model calculation and strong coupling check}
\label{sec:matrix}
The integrated correlator $\mathcal{I}$ can also be obtained using matrix-model techniques. This derivation was presented in \cite{Pufu:2023vwo} and an alternative derivation based on the use of recursion relations and Bessel kernels was discussed in \cite{Billo:2023ncz}.
Here we briefly summarize the latter derivation, which may be useful also for generalizations to massive deformations of $\mathcal{N}=2$ theories, and then analyze the result.

Using supersymmetric localization, the $\mathcal{N}=2^*$ SYM theory on a 4-sphere can be described by a matrix model \cite{Pestun:2007rz} whose partition function, neglecting instanton contributions that are suppressed in the large-$N$ fixed-$\lambda$ limit, can be written as 
\begin{align}
   \cZ(m)=\int \!da~\rme^{-\tr\,a^2}\,
   \rme^{-S_{\rm{int}}(a,\lambda, m)}~. 
\label{Z2*}
\end{align}
Here $a$ is a Hermitian matrix in the fundamental representation of SU($N$), and
\begin{align}
S_{\mathrm{int}}(a,\lambda, m)=-\frac{m^2}{2} \bigg[\sum_{\ell=1}^\infty\sum_{n=0}^{2\ell}(-1)^{n+\ell}\frac{(2\ell+1)!}{n! (2\ell-n)!}\,\zeta_{2\ell+1}\left(\frac{\lambda}{8\pi^2N}\right)^\ell\tr\,a^{2\ell-n}\,\tr\,a^{n}\bigg]+
O(m^4)
\end{align}
where $\zeta_k$ are the Riemann-$\zeta$ values $\zeta(k)$.

The matrix-model counterpart of the vacuum expectation value of the Wilson loop (\ref{vevW}) is given by
\begin{align}
    \cW(m)=\frac{1}{\cZ(m)}\,\int\!
    da~\mathbf{W}(a,\lambda)\,\,\rme^{-\tr a^2}\,\rme^{-S_{\mathrm{int}}(a,\lambda, m)}~,
    \label{eq:matrwl}
\end{align}
where \cite{Pestun:2007rz}
\begin{equation}
\mathbf{W}(a,\lambda)=\frac{1}{N}\,\tr\,\exp\left(\sqrt{\frac{\lambda}{2N}}\,a\right)
~.
\label{WL}
\end{equation}
Thus, the integrated correlator $\cI$ can be obtained by differentiating twice the logarithm of \eqref{eq:matrwl}:
\begin{align}
    \cI&=\partial_m^2\log \cW(m)\Big|_{m=0}=
    \frac{\vev{ \mathbf{W}(a,\lambda)\,S_{\mathrm{int}}(a,\lambda, m)}_0-\vev{ \mathbf{W}(a,\lambda)}_0\,\vev{ S_{\mathrm{int}}(a,\lambda, m)}_0}{\vev{\mathbf{W}(a,\lambda)}_0}~,
    \label{Imatrix}
\end{align}
where the notation $\vev{~}_0$ represents vacuum
expectation values computed in the free Gaussian matrix-model.

In the large-$N$ limit, it is possible to compute the right-hand side of (\ref{Imatrix}) by exploiting the techniques developed in \cite{Beccaria:2021hvt,Billo:2021rdb,Billo:2022gmq,Billo:2022xas,Billo:2022fnb,Billo:2022lrv} and to find the following explicit expression that encodes the exact dependence on the t'Hooft coupling:
\begin{equation}
\label{eq:Yisa}
\cI= \frac{\lambda}{I_1(\sqrt{\lambda)}}
\int_0^\infty\! \frac{dx}{x}~\chi\left(\frac{2\pi x}{\sqrt{\lambda}}\right)\,\frac{\sqrt{\lambda}\,I_0(\sqrt{\lambda})\,J_1(x)^2-x\,I_1(\sqrt{\lambda})\,J_0(x)\,J_1(x)}{x^2+\lambda}+O(1/N^2)~.
\end{equation}
where
\begin{equation}
    \chi(x)=\frac{\big(x/2\big)^2}{\sinh^2\!\left(x/2\right)}~.
\end{equation}
Eq. (\ref{eq:Yisa}) agrees with the results of \cite{Russo:2013kea} 
(see also \cite{Pufu:2023vwo})\,%
\footnote{Notice that in \cite{Russo:2013kea} $\cI$ is not normalized with respect to the vacuum expectation value $\langle \mathbf{W}(a,\lambda)\rangle_0$, and that we differ by an overall factor of 1/2 both with respect to \cite{Russo:2013kea} and \cite{Pufu:2023vwo}.}. 

Expanding (\ref{eq:Yisa}) for small values of $\lambda$, it is straightforward to obtain the perturbative expansion of $\mathcal{I}$, whose first few terms are
\begin{equation}
    \mathcal{I}~\underset{\lambda \rightarrow 0}{\sim}~\frac{3\,\zeta_3}{32\pi^2}\,\lambda^2-\Big(\frac{\zeta_3}{256\pi^2}+\frac{25\,\zeta_5}{256\pi^4}\Big)\lambda^3+\Big(\frac{\zeta_3}{4096 \pi^2}+\frac{15\, \zeta_5}{4096 \pi^4}+\frac{735\,\zeta_7}{8192 \pi^6}\Big)\lambda^4+O(\lambda^5)+\dots
    \label{eq:Ipert}
\end{equation}
where the ellipses stand for sub-leading terms in the large-$N$
expansion.
On the other hand, using the asymptotic behavior of the Bessel functions for large values of their arguments, we can compute from (\ref{eq:Yisa}) the strong-coupling expansion of $\mathcal{I}$ and get
\begin{equation}
    \mathcal{I}~\underset{\lambda \rightarrow \infty}{\sim}~
    \frac{\sqrt{\lambda}}{2}+\Big(\frac{1}{4}-\frac{\pi^2}{6}\Big)+
    O(\lambda^{-1/2})+\dots~.
    \label{eq:Istrong}
\end{equation}

\subsection{Matching with the matrix-model results at strong coupling}

The expansions (\ref{eq:Ipert}) and (\ref{eq:Istrong}) should be reproduced from the gauge theory side by computing the integral in (\ref{Afin}) with the weak and strong-coupling expression of the function $\widehat{F}(\rho,\eta)$ appearing in the 2-point function of the moment-map operator $J$ with a Wilson line.
At weak coupling this correlator has been computed explicitly only for specific configurations (see for example \cite{Barrat:2020vch} where the collinear limit is considered) and thus there are no available general expressions for $\widehat{F}(\rho,\eta)$ that can be used to check the integrated correlator (\ref{Afin}) in this regime. On the contrary, at strong coupling this correlator has been computed in \cite{Barrat:2021yvp} at leading order in the large-$N$ expansion using bootstrap techniques. In particular, the function $\widehat{F}(\rho,\eta)$ can be deduced from the quantity $F_0$ given in the first line of eq. (3.33) in \cite{Barrat:2021yvp}. In our normalizations, we have
\begin{align}
    \widehat{F}(\rho,\eta)
    &~\underset{\lambda \rightarrow 0}{\sim}~\frac{\sqrt{\lambda}}{32\pi^4}\Bigg[-\frac{1}{2}\frac{z \overbar{z}}{(1-z)(1-\overbar{z})}\Bigg(\frac{1+z\overbar{z}}{(1-z\overbar{z})^2}+\frac{2z \overbar{z}\log z \overbar{z}}{(1-z\overbar{z})^3}\Bigg)\Bigg]+O(\lambda^0)
    \label{F0}
\end{align}
where
\begin{equation}
    z=\rho\,\mathrm{e}^{i\theta}\quad,\quad
    \overbar{z}=\rho\,\mathrm{e}^{-i\theta}\, .
\end{equation}
More explicitly, the strong-coupling expression of $\hat{F}$ at leading order is
\begin{align}
    \widehat{F}(\rho,\eta)=\frac{\sqrt{\lambda}}{64\pi^4}\,\frac{\rho ^2-\rho ^6+4 \rho ^4 \log \rho}{\left(\rho ^2-1\right)^3
   \left(1+\rho^2-2 \eta  \rho\right)}~.
\end{align}
Using this in (\ref{Afin}) and performing the integrals,
we find
\begin{align}
    \mathcal{I}&=128\pi^4\!\!\int_0^1\!\!d\rho\int_{-1}^{+1}\!\!d\eta\,\,\frac{\rho^2-1}{4\rho^2}\,\hat{F}(\rho,\eta)\notag\\
    &=\frac{\sqrt{\lambda}}{2}\!\int_0^1\!\!d\rho\int_{-1}^{+1}\!\!d\eta\,\,\frac{1-\rho^4+4 \rho^2 
    \log \rho}{\left(\rho ^2-1\right)^2
   \left(1+\rho^2-2 \eta  \rho\right)}=\frac{\sqrt{\lambda}}{2}
\end{align}
which is exactly the leading strong-coupling result given in (\ref{eq:Istrong}). This successful check is a strong indication of the validity of the integration measure we have found and of the integral constraint (\ref{Afin}).

\vskip 1cm
\noindent {\large {\bf Acknowledgments}}
\vskip 0.2cm
We would like to thank Lorenzo Bianchi, Maddalena Ferragatta, Alessandro Georgoudis and Marco Meineri for many useful discussions. We thank in particular Lorenzo Bianchi and Maddalena Ferragatta for having drawn our attention to the possible presence also of parity-odd terms in the current/scalar defect correlators.\\
This research is partially supported by the MUR PRIN contract 2020KR4KN2 ``String Theory as a bridge between Gauge Theories and Quantum Gravity'' and by
the INFN project ST\&FI
``String Theory \& Fundamental Interactions''.  F.G. is supported by a STFC Consolidated Grant, ST$\backslash$T000686$\backslash$1 ``Amplitudes, strings \& duality". 
\vskip 1cm

\appendix
\section{Parity odd terms}
\label{app_odd}
We follow the embedding formalism, which provides a realization of conformal transformations as linear coordinate transformations on the light-cone of $ \mathbb{R}^{1,d+1}$. A detailed derivation of this formalism in presence of a defect can be found in \cite{Billo:2016cpy,Lauria:2018klo}, in this appendix we report the results that are useful in the main text.

The spacetime embedding coordinates in $ \mathbb{R}^{1,d+1} $ are realised as
\begin{equation}
P^M = \left(\frac{1+x^2}{2},x^{\mu},\frac{1-x^2}{2}\right),
\end{equation}
defined on the light cone $P^2=0$. Analogously we can define spinning operators in the embedding space by introducing auxiliary vectors $Z_M$:
\begin{equation}
Z_M=(x\cdot z, x^\mu,-x\cdot z)~,
\end{equation}
which again must satisfy $Z^2=0$.
For our purposes we need the embedding space version of a spin one operator:
\begin{equation}\label{4.2.5}
\cO(P,Z) = Z^{M}\cO_M(P)~,
\end{equation}
which satisfies $P\cdot Z =0$.

In presence of a line defect in 4d, it is natural to split the coordinates in two sets $M=(A,I)$, distinguishing the ``parallel" directions denoted by $A,B= 0,1,2$ indices, and ``orthogonal" directions denoted by $I,J=3,4,5$. The broken conformal group SO$(1,2)$ and SO$(3)$ act on these sets separately:
The residual conformal symmetry is still linearly realised, and one can define two different distances, in the parallel and transverse directions:
\begin{equation}\label{eq:defectDist}
P\bullet Q=P^A\eta_{AB}Q^B \hspace{1cm} P\circ Q= P^I\delta_{IJ} Q^J
\end{equation}
Since bulk insertions still satisfy the conditions $P^2=Z^2=Z\cdot P=0$, only a subset of the scalar products \eqref{eq:defectDist} is independent
\begin{equation}
P \bullet P = - P \circ P~, ~~~~~ Z \bullet Z=-Z\circ Z~, ~~~~~ Z \bullet P = -Z \circ P~.
\end{equation}
As a first example we write the bulk two-point functions of two scalar operators with dimensions $\Delta_1$ and $\Delta_2$:
\begin{equation}
    \vev{O(P_1)O(P_2)}_W = \frac{f(\rho,\eta)}{(P_1\circ P_1)^{\Delta_1/2}(P_2\circ P_2)^{\Delta_2/2}}~,
\end{equation}
which can be used to fix scalar two-point functions in physical space as \eqref{WJJWKK}, \eqref{Avsa} and \eqref{eq:TT_QQ} in the main text.

We then consider bulk two-point functions in presence of spin-one operators. We start from the current-current two-point function, which is fixed as follows:
\begin{align}\label{eq:JJ_embed}
    \vev{J(P_1,Z_1)J(P_2,Z_2)}_W = \frac{1}{(P_1\circ P_1)^{\Delta_1/2}(P_2\circ P_2)^{\Delta_2/2}}\sum_k f_k(\rho,\eta)Q_k(P_1,Z_1,P_2,Z_2)~,
\end{align}
where $u,v$ are the usual cross ratios and $k$ counts the possible tensorial structures allowed by conformal symmetry. For the correlator of two spin one currents there are 6 possible tensorial structures $Q_k$, as listed in \cite{Billo:2016cpy,Lauria:2018klo}, whose expressions in the physical space generate the the correlators $\vev{j^m j^n}$ displayed in \eqref{eq:jj_general}. This correlator cannot be modified by any additional parity-odd structure, so the results in \cite{Lauria:2018klo} are compatible with \cite{Herzog:2020bqw} and the Ward identity computation in \cite{Billo:2023ncz}.

The two-point function of a current with a scalar operator
\begin{align}\label{JO_embedding}
\langle \cO_{\Delta_1}(P_1,Z_1)\cO_{\Delta_2}(P_2)\rangle_W
\end{align}
is fixed analogously to \eqref{eq:JJ_embed}, where, as prescribed by \cite{Lauria:2018klo}, there are only two allowed conformal structures $Q_k$. Hence, their expressions in physical space generate \eqref{eq:jT_general} and \eqref{eq:jQ_general} in the main text. However, as mentioned in \cite{Billo:2016cpy} (see in particular their section 3.4) one must take into account possible parity-odd structures. Indeed, in presence of a defect the residual conformal symmetry group connected to the identity does not relate parity transformations applied separately on parallel and orthogonal directions. For a line defect and for observables like \eqref{JO_embedding}, it is then possible to saturate the vectors $(P_1,Z_1,P_2)$ with the appropriate parallel and orthogonal epsilon tensors \footnote{Similar cases were encountered for bulk one-point function in presence of line and surface defects in \cite{Bianchi:2019dlw,Bianchi:2019sxz}.}
\begin{equation}
\epsilon_{A_1A_2A_3}~,~~~ A_a=0,1,2~, \hspace{1cm}\epsilon_{I_1I_2I_3}~,~~~ I_i=3,4,5~.
\end{equation}
We can think of $\epsilon_{A_1A_2A_3}$ as the time reversal structure and the $\epsilon_{I_1I_2I_3}$ as the spacial parity reversal structure.
Therefore in this case we must add to the structures listed in \cite{Lauria:2018klo} the following additional functions:
\begin{align}\label{4.2.38}
\begin{split}
\langle \cO_{\Delta_1}(P_1,Z_1)\cO_{\Delta_2}(P_2)\rangle_W &= \frac{1}{(P_1\circ P_1)^{\frac{\Delta_1+1}{2}}(P_2\circ P_2)^{\frac{\Delta_2+1}{2}}}\Big(s(\rho,\eta)~\epsilon_{I_1I_2I_3}P_1^{I_1}~P_2^{I_2}~Z_1^{I_3}\\
&~~~~-2s^\prime(\rho,\eta) \epsilon_{A_1A_2A_3}P_1^{A_1}~P_2^{A_2}~Z_1^{A_3}\Big)~.
\end{split}
\end{align}
where $s(\rho,\eta)$ and $s^\prime(\rho,\eta)$ are two additional functions of the cross ratios. Scalar operators transforming nontrivially under time/spacial parity reversal can allow for these epsilon structures. 

Therefore, we can map these additional structures to the 4d spacetime using the Poincar\`e section. The trasverse $\epsilon_{I_1I_2I_3}$ gives:
\begin{align}
\vev{j_m (x_1) O_{\Delta_2}(x_2)}_W = s\, \frac{\epsilon_{mnp} x_1^n x_2^p}{r_1^{\Delta_1+1}r_2^{\Delta_2+1}}~,~~~~~~\vev{j_4 (x_1) O_{\Delta_2}(x_2)}_W=0~,
\end{align}
whereas the parallel $\epsilon_{A_1A_2A_3}$ gives:
\begin{align}
\vev{j_m (x_1) \tilde O_{\Delta_2}(x_2)}_W = 2s^\prime \frac{t_{12}x_{1m}}{r_1^{\Delta_1+1}r_2^{\Delta_2+1}}~,~~~~~~\vev{j_4 (x_1) \tilde O_{\Delta_2}(x_2)}_W = s^\prime\,\frac{t_{12}^2-r_1^2+r_2^2}{r_1^{\Delta_1+1}r_2^{\Delta_2+1}} ~,
\end{align}
which give rise to the formulas \eqref{Todd} and \eqref{Qodd} in the main text.

\bibliography{biblio}

\providecommand{\href}[2]{#2}\begingroup\raggedright\begin{thebibliography}{10}

\bibitem{Binder:2019jwn}
D.~J. Binder, S.~M. Chester, S.~S. Pufu, and Y.~Wang, \emph{{$ \mathcal{N} $ =
  4 Super-Yang-Mills correlators at strong coupling from string theory and
  localization}}, \href{http://dx.doi.org/10.1007/JHEP12(2019)119}{JHEP {\bf
  12} (2019)  119}, \href{http://arxiv.org/abs/1902.06263}{{\tt
  arXiv:1902.06263 [hep-th]}}.

\bibitem{Chester:2019jas}
S.~M. Chester, M.~B. Green, S.~S. Pufu, Y.~Wang, and C.~Wen, \emph{{Modular
  invariance in superstring theory from $ \mathcal{N} $ = 4 super-Yang-Mills}},
  \href{http://dx.doi.org/10.1007/JHEP11(2020)016}{JHEP {\bf 11} (2020)  016},
  \href{http://arxiv.org/abs/1912.13365}{{\tt arXiv:1912.13365 [hep-th]}}.

\bibitem{Chester:2020dja}
S.~M. Chester and S.~S. Pufu, \emph{{Far beyond the planar limit in
  strongly-coupled $ \mathcal{N} $ = 4 SYM}},
  \href{http://dx.doi.org/10.1007/JHEP01(2021)103}{JHEP {\bf 01} (2021)  103},
  \href{http://arxiv.org/abs/2003.08412}{{\tt arXiv:2003.08412 [hep-th]}}.

\bibitem{Chester:2020vyz}
S.~M. Chester, M.~B. Green, S.~S. Pufu, Y.~Wang, and C.~Wen, \emph{{New modular
  invariants in $ \mathcal{N} $ = 4 Super-Yang-Mills theory}},
  \href{http://dx.doi.org/10.1007/JHEP04(2021)212}{JHEP {\bf 04} (2021)  212},
  \href{http://arxiv.org/abs/2008.02713}{{\tt arXiv:2008.02713 [hep-th]}}.

\bibitem{Dorigoni:2021bvj}
D.~Dorigoni, M.~B. Green, and C.~Wen, \emph{{Novel Representation of an
  Integrated Correlator in $\mathcal N$ = 4 Supersymmetric Yang-Mills Theory}},
  \href{http://dx.doi.org/10.1103/PhysRevLett.126.161601}{Phys. Rev. Lett. {\bf
  126} (2021) no.~16, 161601}, \href{http://arxiv.org/abs/2102.08305}{{\tt
  arXiv:2102.08305 [hep-th]}}.

\bibitem{Dorigoni:2021guq}
D.~Dorigoni, M.~B. Green, and C.~Wen, \emph{{Exact properties of an integrated
  correlator in $ \mathcal{N} $ = 4 SU(N) SYM}},
  \href{http://dx.doi.org/10.1007/JHEP05(2021)089}{JHEP {\bf 05} (2021)  089},
  \href{http://arxiv.org/abs/2102.09537}{{\tt arXiv:2102.09537 [hep-th]}}.

\bibitem{Dorigoni:2022iem}
D.~Dorigoni, M.~B. Green, and C.~Wen, \emph{{The SAGEX review on scattering
  amplitudes Chapter 10: Selected topics on modular covariance of type IIB
  string amplitudes and their~~supersymmetric Yang\textendash{}Mills duals}},
  \href{http://dx.doi.org/10.1088/1751-8121/ac9263}{J. Phys. A {\bf 55} (2022)
  no.~44, 443011}, \href{http://arxiv.org/abs/2203.13021}{{\tt arXiv:2203.13021
  [hep-th]}}.

\bibitem{Pufu:2023vwo}
S.~S. Pufu, V.~A. Rodriguez, and Y.~Wang, \emph{{Scattering From
  $(p,q)$-Strings in $\text{AdS}_5 \times \text{S}^5$}},
  \href{http://arxiv.org/abs/2305.08297}{{\tt arXiv:2305.08297 [hep-th]}}.

\bibitem{Billo:2023ncz}
M.~Billo, F.~Galvagno, M.~Frau, and A.~Lerda, \emph{{Integrated correlators
  with a Wilson line in $ \mathcal{N} $ = 4 SYM}},
  \href{http://dx.doi.org/10.1007/JHEP12(2023)047}{JHEP {\bf 12} (2023)  047},
  \href{http://arxiv.org/abs/2308.16575}{{\tt arXiv:2308.16575 [hep-th]}}.

\bibitem{Pestun:2007rz}
V.~Pestun, \emph{{Localization of gauge theory on a four-sphere and
  supersymmetric Wilson loops}},
  \href{http://dx.doi.org/10.1007/s00220-012-1485-0}{Commun. Math. Phys. {\bf
  313} (2012)  71--129}, \href{http://arxiv.org/abs/0712.2824}{{\tt
  arXiv:0712.2824 [hep-th]}}.

\bibitem{Buchel:2013id}
A.~Buchel, J.~G. Russo, and K.~Zarembo, \emph{{Rigorous Test of Non-conformal
  Holography: Wilson Loops in N=2* Theory}},
  \href{http://dx.doi.org/10.1007/JHEP03(2013)062}{JHEP {\bf 03} (2013)  062},
  \href{http://arxiv.org/abs/1301.1597}{{\tt arXiv:1301.1597 [hep-th]}}.

\bibitem{Russo:2013kea}
J.~G. Russo and K.~Zarembo, \emph{{Massive N=2 Gauge Theories at Large N}},
  \href{http://dx.doi.org/10.1007/JHEP11(2013)130}{JHEP {\bf 11} (2013)  130},
  \href{http://arxiv.org/abs/1309.1004}{{\tt arXiv:1309.1004 [hep-th]}}.

\bibitem{Lauria:2018klo}
E.~Lauria, M.~Meineri, and E.~Trevisani, \emph{{Spinning operators and defects
  in conformal field theory}},
  \href{http://dx.doi.org/10.1007/JHEP08(2019)066}{JHEP {\bf 08} (2019)  066},
  \href{http://arxiv.org/abs/1807.02522}{{\tt arXiv:1807.02522 [hep-th]}}.

\bibitem{Herzog:2020bqw}
C.~P. Herzog and A.~Shrestha, \emph{{Two point functions in defect CFTs}},
  \href{http://dx.doi.org/10.1007/JHEP04(2021)226}{JHEP {\bf 04} (2021)  226},
  \href{http://arxiv.org/abs/2010.04995}{{\tt arXiv:2010.04995 [hep-th]}}.

\bibitem{Barrat:2021yvp}
J.~Barrat, A.~Gimenez-Grau, and P.~Liendo, \emph{{Bootstrapping holographic
  defect correlators in $ \mathcal{N} $ = 4 super Yang-Mills}},
  \href{http://dx.doi.org/10.1007/JHEP04(2022)093}{JHEP {\bf 04} (2022)  093},
  \href{http://arxiv.org/abs/2108.13432}{{\tt arXiv:2108.13432 [hep-th]}}.

\bibitem{Gimenez-Grau:2023fcy}
A.~Gimenez-Grau, \emph{{The Witten Diagram Bootstrap for Holographic Defects}},
  \href{http://arxiv.org/abs/2306.11896}{{\tt arXiv:2306.11896 [hep-th]}}.

\bibitem{Barrat:2020vch}
J.~Barrat, P.~Liendo, and J.~Plefka, \emph{{Two-point correlator of chiral
  primary operators with a Wilson line defect in $ \mathcal{N} $ = 4 SYM}},
  \href{http://dx.doi.org/10.1007/JHEP05(2021)195}{JHEP {\bf 05} (2021)  195},
  \href{http://arxiv.org/abs/2011.04678}{{\tt arXiv:2011.04678 [hep-th]}}.

\bibitem{DOPW}
R.~Dempsey, B.~Offertaler, S.~S. Pufu, and Y.~Wang, \emph{{Global Symmetry and
  Integral Constraint on Superconformal Lines in Four Dimensions}}, to appear.

\bibitem{Dolan:2001tt}
F.~A. Dolan and H.~Osborn, \emph{{Superconformal symmetry, correlation
  functions and the operator product expansion}},
  \href{http://dx.doi.org/10.1016/S0550-3213(02)00096-2}{Nucl. Phys. B {\bf
  629} (2002)  3--73}, \href{http://arxiv.org/abs/hep-th/0112251}{{\tt
  arXiv:hep-th/0112251}}.

\bibitem{Billo:2016cpy}
M.~Billo, V.~Gon\c{c}alves, E.~Lauria, and M.~Meineri, \emph{{Defects in
  conformal field theory}},
  \href{http://dx.doi.org/10.1007/JHEP04(2016)091}{JHEP {\bf 04} (2016)  091},
  \href{http://arxiv.org/abs/1601.02883}{{\tt arXiv:1601.02883 [hep-th]}}.

\bibitem{Buchbinder:2012vr}
E.~I. Buchbinder and A.~A. Tseytlin, \emph{{Correlation function of circular
  Wilson loop with two local operators and conformal invariance}},
  \href{http://dx.doi.org/10.1103/PhysRevD.87.026006}{Phys. Rev. D {\bf 87}
  (2013) no.~2, 026006}, \href{http://arxiv.org/abs/1208.5138}{{\tt
  arXiv:1208.5138 [hep-th]}}.

\bibitem{Beccaria:2021hvt}
M.~Beccaria, M.~Billo, M.~Frau, A.~Lerda, and A.~Pini, \emph{{Exact results in
  a $ \mathcal{N} $ = 2 superconformal gauge theory at strong coupling}},
  \href{http://dx.doi.org/10.1007/JHEP07(2021)185}{JHEP {\bf 07} (2021)  185},
  \href{http://arxiv.org/abs/2105.15113}{{\tt arXiv:2105.15113 [hep-th]}}.

\bibitem{Billo:2021rdb}
M.~Billo, M.~Frau, F.~Galvagno, A.~Lerda, and A.~Pini, \emph{{Strong-coupling
  results for $ \mathcal{N} $ = 2 superconformal quivers and holography}},
  \href{http://dx.doi.org/10.1007/JHEP10(2021)161}{JHEP {\bf 10} (2021)  161},
  \href{http://arxiv.org/abs/2109.00559}{{\tt arXiv:2109.00559 [hep-th]}}.

\bibitem{Billo:2022gmq}
M.~Billo, M.~Frau, A.~Lerda, A.~Pini, and P.~Vallarino, \emph{{Structure
  constants in $\mathcal{N}=2$ superconformal quiver theories at strong
  coupling and holography}},
  \href{http://dx.doi.org/10.1103/PhysRevLett.129.031602}{Phys. Rev. Lett. {\bf
  129} (2022) no.~3, 031602}, \href{http://arxiv.org/abs/2206.13582}{{\tt
  arXiv:2206.13582 [hep-th]}}.

\bibitem{Billo:2022xas}
M.~Billo, M.~Frau, A.~Lerda, A.~Pini, and P.~Vallarino, \emph{{Three-point
  functions in a $ \mathcal{N} $ = 2 superconformal gauge theory and their
  strong-coupling limit}},
  \href{http://dx.doi.org/10.1007/JHEP08(2022)199}{JHEP {\bf 08} (2022)  199},
  \href{http://arxiv.org/abs/2202.06990}{{\tt arXiv:2202.06990 [hep-th]}}.

\bibitem{Billo:2022fnb}
M.~Billo, M.~Frau, A.~Lerda, A.~Pini, and P.~Vallarino, \emph{{Localization vs
  holography in 4d $\mathcal{N} $ = 2 quiver theories}},
  \href{http://dx.doi.org/10.1007/JHEP10(2022)020}{JHEP {\bf 10} (2022)  020},
  \href{http://arxiv.org/abs/2207.08846}{{\tt arXiv:2207.08846 [hep-th]}}.

\bibitem{Billo:2022lrv}
M.~Billo, M.~Frau, A.~Lerda, A.~Pini, and P.~Vallarino, \emph{{Strong coupling
  expansions in $ \mathcal{N} $ = 2 quiver gauge theories}},
  \href{http://dx.doi.org/10.1007/JHEP01(2023)119}{JHEP {\bf 01} (2023)  119},
  \href{http://arxiv.org/abs/2211.11795}{{\tt arXiv:2211.11795 [hep-th]}}.

\bibitem{Bianchi:2019dlw}
L.~Bianchi, M.~Billo, F.~Galvagno, and A.~Lerda, \emph{{Emitted Radiation and
  Geometry}}, \href{http://dx.doi.org/10.1007/JHEP01(2020)075}{JHEP {\bf 01}
  (2020)  075},
\href{http://arxiv.org/abs/1910.06332}{{\tt arXiv:1910.06332 [hep-th]}}.

\bibitem{Bianchi:2019sxz}
L.~Bianchi and M.~Lemos, \emph{{Superconformal surfaces in four dimensions}},
  \href{http://dx.doi.org/10.1007/JHEP06(2020)056}{JHEP {\bf 06} (2020)  056},
  \href{http://arxiv.org/abs/1911.05082}{{\tt arXiv:1911.05082 [hep-th]}}.

\end{thebibliography}\endgroup

\end{document}